\documentclass{article}
\usepackage[T1]{fontenc} \usepackage[utf8]{inputenc} \usepackage{ismir,amsmath,cite,url}
\usepackage{amssymb}
\usepackage{amsthm}
\theoremstyle{definition}
\usepackage[export]{adjustbox}
\newtheorem{definition}{Definition}[section]
\usepackage{graphicx}
\usepackage{color}
\usepackage{cleveref}
\usepackage{amssymb}
\usepackage{booktabs}
\usepackage{subcaption}

\usepackage[normalem]{ulem}
\usepackage{soul}
\usepackage{lineno}

\title{Exploring Sampling Techniques for Generating Melodies with a Transformer Language Model}

\oneauthor {Mathias Rose Bjare$^1$ \hspace{1.5cm} Stefan Lattner$^2$ \hspace{1.5cm}  Gerhard Widmer $^{1,3}$} {$^1$ Institute of Computational Perception, Johannes Kepler University Linz, Austria\\ $^2$ Sony Computer Science Laboratories (CSL), Paris, France\\ $^3$ LIT AI Lab, Linz Institute of Technology, Austria\\ {\tt 
mathias.bjare@jku.at}}

\def\authorname{M. Bjare, S. Lattner, and G. Widmer}

\begin{document}

\maketitle
\begin{abstract}
Research in natural language processing has demonstrated that the quality of generations from trained autoregressive language models is significantly influenced by the used sampling strategy.
In this study, we investigate the impact of different sampling techniques on musical qualities such as diversity and structure. To accomplish this, we train a high-capacity transformer model on a vast collection of highly-structured Irish folk melodies and analyze the musical qualities of the samples generated using distribution truncation sampling techniques. Specifically, we use nucleus sampling, the recently proposed "typical sampling", and conventional ancestral sampling.
We evaluate the effect of these sampling strategies in two scenarios: optimal circumstances with a well-calibrated model and suboptimal circumstances where we systematically degrade the model's performance. We assess the generated samples using objective and subjective evaluations. We discover that probability truncation techniques may restrict diversity and structural patterns in optimal circumstances, but may also produce more musical samples in suboptimal circumstances.
\end{abstract}
\section{Introduction}\label{sec:introduction}
In recent years, developments in natural language modelling have also accelerated the field of symbolic music generation. In this context, the musical events of a music piece are represented as a sequence of symbols or tokens from a fixed vocabulary, and the goal is to learn to generate new token sequences. 
At present, the autoregressive transformer model \cite{transformer} is the basis of many symbolic music generation models \cite{musictransformer,remi, remip, museformer}.
In this context, a conditional distribution is learned by solving a masked self-prediction task \cite{musictransformer,remi, remip, museformer}, and generation is performed with stochastic sampling techniques, e.g., ancestral sampling, or maximization-based search techniques, e.g., beam search. 

However, the choice of decoding technique has been shown to impact various qualitative features of generated samples substantially. In \cite{nucleus}, the authors showed that generation with \textit{nucleus sampling} yields natural language samples that are more contextualized than those from conventional sampling techniques, and samples of nucleus sampling score higher in human evaluations. More recently, the authors of \cite{typicaldecoding} propose \textit{typical sampling} and show that it reduces degenerate sample generation while exhibiting performance competitive with nucleus sampling. Typical sampling is based on the authors' finding that words in human language are \textit{typical}. More specifically, the authors show that most words of human language are, in fact, not the most likely words (lowest information content (IC)), as measured with a language model, but rather \textit{typical} words, i.e., they have an IC close to the conditional entropy of the language model. Typical sampling explicitly enforces this condition.

We hypothesize that a careful choice of sampling technique could also improve certain aspects of music generated using language models, particularly because in \cite{bjarelattner}, it has been shown that musical events tend to be typical. However, we find that many music generation systems rely on ordinary sampling techniques. In addition, studies on the effect of sampling techniques on musical qualities are limited.

In this work, we study the structural and tonal properties of music generated with different sampling techniques applied to a high-capacity transformer model. Specifically, we measure the IC, long and short-term self-similarities and scale consistency of samples generated with conventional sampling, nucleus sampling, and typical sampling. We test the sampling techniques for a well-calibrated model and for under-calibrated models. We support our findings by performing a listening study.
We conduct our experiments on \textit{The Session} dataset \cite{folkrnnsession}, a large dataset of well-structured monophonic music in the established musical genre of Irish traditional music. We choose this dataset since we expect it to provide suitable conditions for training a well-calibrated model. Our findings suggest that truncation techniques can address inadequacies of models that are not well-fitted to the data.

\section{Background and Related work}
\label{sec:background}
Although maximization-based techniques like beam search work well for directed language generation tasks\footnote{Generation with input sequence conditioning.} (such as machine translation and summarization), beam search has been shown to produce dull and repetitive samples for open-ended language generation tasks\footnote{Generation without input sequence conditioning.} \cite{nucleus}, an effect that can be observed in music generation as well \cite{dieleman2020typicality}.  It is, therefore, more common to use stochastic sampling techniques\footnote{In the context of generative models, ``\textit{sampling techniques}'' could refer to a multitude of aspects in the generative pipeline (e.g., Gibbs sampling in restricted Boltzmann machines \cite{rbm}). In our work, ``\textit{sampling techniques}'', refers to techniques for obtaining samples from a trained language model.} for open-ended generation tasks. The most obvious method is ancestral sampling, where one token at a time is sampled based on the predicted distribution, conditioned on the previously generated tokens. However, it has been shown that truncating the conditional distribution (by setting the probability of specific tokens to zero, followed by renormalising), can lead to better sample quality than the non-truncated variant. An example of distribution truncation is top-$k$ sampling, where all but the $k$ most probable tokens are zeroed. In \cite{gpt2}, the authors showed that top-\textit{k} sampling generates more coherent samples 
than the non-truncated variant. In \cite{nucleus}, it is explained that the quality improvement of top-$k$ sampling is caused by removing unreliably estimated low-probability tokens, and it is found that top-$k$ sampling mitigates the problem. However, it is also shown that top-$k$ sampling is sensitive to the distribution's entropy (see \Cref{sec:typical}), making it hard to select a value of $k$ that fits both high and low certainty conditions. As a solution, they propose \textit{nucleus sampling} that assigns zero probability to the largest set of least probable tokens that together have a probability below a given threshold.
The authors find that the samples produced using the technique are preferred by humans over other sampling techniques. Nucleus sampling has been used in music generation in \cite{nucleusmusic,nucleusmusic2,nucleus3}, but its effects are difficult to quantify without comparisons to the non-truncated case. Although nucleus sampling mitigates the problem of poorly estimated low-probability tokens, it does not prevent generating degenerated repetitive sequences caused by low entropy distributions (see \Cref{sec:methods}). As a solution, in \cite{typicaldecoding}, the authors propose \textit{typical sampling} and show that this technique prevents degenerated sample generation. 
\section{Ancestral sampling}
\label{sec:methods}
Let $p\left(x_{t}\vert x_{<t}\right)$ be the conditional probability of a symbol $x_{t}$ given previously observed symbols $x_{<t}$ (i.e., the context) and let $q$ be a model fitted to $p$, e.g., a neural network fitted via likelihood maximization. 
Given a model $q$, ancestral sampling samples one token at a time using $x_{0}\sim q(\cdot), x_{1}\sim q(\cdot| x_{0}),..., x_{t} \sim q(\cdot | x_{<t})$.

\subsection{Distribution truncation sampling techniques}
In distribution truncation, a truncated distribution $\widetilde{q}$ is obtained by zeroing the probability of a subset of tokens and renormalising the resulting distribution. Formally, $\widetilde{q}$ is defined by
\begin{equation}
    \widetilde{q}(x_t | x_{<t}) = 
    \begin{cases}
    \frac{q(x_t | x_{<t})}{\sum_{v \in V} q(v | x_{<t})} &\text{if } x_{t} \in V\\
    0 & \text{otherwise}
    \end{cases},
    \end{equation}
where $V$ is the set of tokens with nonzero probability in $\widetilde{q}$. 
For the remainder of this article, we use `\textit{conventional sampling}' to denote sampling from untruncated distributions.

\subsection{Nucleus sampling}
In nucleus sampling, $V$ is defined as the smallest set such that 
\begin{equation}
    \sum_{v\in V} q(v \vert x_{<t}) \geq \tau,
\end{equation}
where $\tau$ is a constant determining the number of tokens to be removed. 

\subsection{Typical sampling}
\label{sec:typical}
\begin{figure}
    \centering
    \includegraphics[scale=.45,trim={1.8cm 0 0 0}, clip]{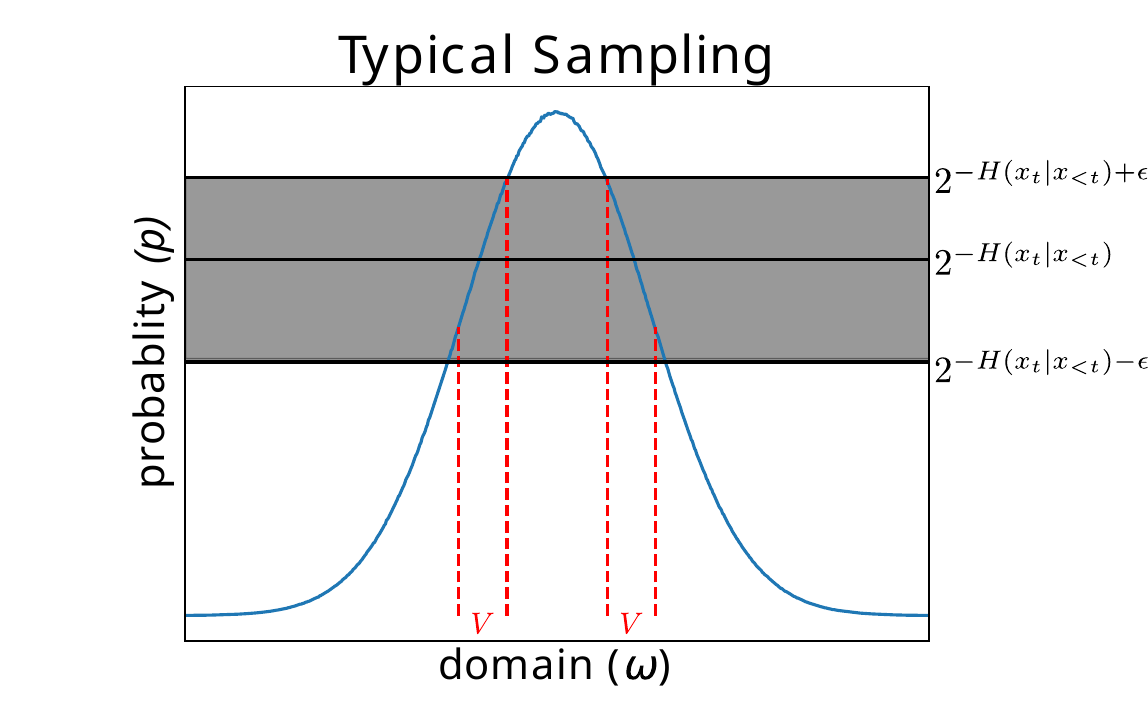}
    \caption{
    In typical sampling, a probability band around the entropy (dark-grey) defines the set $V$ of tokens with non-zero probabilities in the truncated distribution.}
    \label{fig:typ}
\end{figure}

In typical sampling \cite{typicaldecoding}, $V$ is defined in terms of the token information content described below. 
\begin{definition}[Conditional information content]
The conditional \textit{information content} (IC) is given by
\begin{equation}
IC\left(x_{t}\vert x_{<t}\right) = -\log{q\left(x_{t}\vert x_{<t}\right)}.        
\end{equation}
\end{definition}
In computational music perception, IC has been used to model how surprising a musical event is given the musical context \cite{meyer,idyom, tismir}.
\begin{definition}[Conditional entropy]
    The conditional entropy is the expected conditional information content
\begin{equation}
    H\left(x_{t}\vert x_{<t}\right) = \mathbb{E}_{x_{t} \sim q(\cdot\vert x_{<t})}\left[IC\left(x_{t}\vert x_{<t}\right)\right].
\end{equation}
\end{definition}
The entropy of a distribution explains how confident a model is. It ranges from $0$ to $\log{n}$ where $n$ is the number of symbols in the vocabulary, with $0$ indicating that the distribution is deterministic and $\log{n}$ indicating that the distribution is uniformly random.
In typical sampling, the probabilities of tokens with the highest deviation of information from the entropy
\begin{align}
        \left| H\left(x_{t}\vert x_{<t}\right) - IC\left(x_{t}\vert x_{<t}\right)\right|  \label{eq:typ_evt}
\end{align} are set to zero.
More precisely, let $U = v_1,v_2,...,v_{n}$ be an ascending ordering of the vocabulary in accordance to \Cref{eq:typ_evt}. Then $V$ is defined as the smallest prefix of $U$ such that $q(U\vert x_{<t}) \geq \tau$. \Cref{eq:typ_evt} implies that $V$ is restricted by a band around the entropy as shown in \Cref{fig:typ}. Therefore, also the most likely token under $q$ can have zero probability in $\tilde{q}$. The authors of \cite{typicaldecoding} note that this property, however, lowers the number of degenerately repetitive samples, as opposed to nucleus sampling, without degrading preference in human evaluations.

\section{Experiments}\label{sec:exp}
In this section, we describe the setup for both our objective and subjective experiments, the used data, training details, degradation scenarios, and generation details, as well as objective and subjective evaluation.

\subsection{Data}

Our experiments are performed on monophonic symbolic music. 
Specifically, we use the midi-encoded version of the \textit{The Session} dataset \cite{folkrnnsession}, consisting of $45,849$ traditional Irish folk tunes originally encoded in ABC notation.
We discard the $5\%$ longest sequences to lower the computational footprint of the autoregressive transformer model, and
partition the dataset in training, validation, and test sets with proportions 10/12, 1/12, and 1/12, respectively. All analyses will be performed on the test set, while our generative models will be trained and optimized on the training and validation sets, respectively. The dataset contains tunes with the same name, corresponding to different versions of the same tune. We ensure that tunes with the same name appear in exactly one of the three sets. 

We tokenize the sequences using a modified version of the popular REMI representation \cite{remi}. REMI serializes a score bar-wise from left to right. A bar is serialized as a sequence of tokens starting with a bar-delimiter token followed by a serialization of the notes within that bar. Each note is serialized as three tokens indicating the onset within the bar, the pitch and the duration, in that order. The position and duration tokens are quantized to 1/12th of a beat. Contrary to the original REMI implementation, we omit velocity, tempo change and chord symbols, since these are not encoded in the original ABC files either. Similar to \cite{remip}, we extend the REMI representation with time-signature tokens inserted immediately after the bar token. We base our tokenization implementation on a modified version of the REMI python implementation in MidiTok \cite{miditok2021}.

\subsection{Training}
We train a 21-layered Transformer decoder model \cite{attention} with relative attention \cite{relativeattention, musictransformer} in a self-supervised prediction task.
We train the model using Adam optimization \cite{adam} with a learning rate of $10^{-4}$ until no improvement takes place on the validation set during 10 subsequent epochs. The used batch size is 16, and the input sequence length is 512 tokens. Sequences shorter than 512 tokens are zero-padded. The negative log-likelihood (NLL) on the test dataset is measured to be $NLL=0.30$, which is similar to the result of a recent transformer-based model trained on the same dataset \cite{tismir}.
We thus call this a \textit{well-calibrated model}.

\subsection{Model Degradation}
In addition to the well-calibrated model, we consider two \textit{under-calibrated models}, which we achieve by intentionally degrading the well-calibrated model. 
For our first degradation, we scale the logits vector $h$ of the transformer softmax output distribution, i.e., 
\begin{equation}
    q(x_t \vert x_{<t}) = \text{Softmax}(h/r),
\end{equation}
where $r>1.0$ is a temperature scale. This degradation increases the distribution's entropy (uncertainty) while keeping the relative ordering of the probabilities the same. Using temperature scaling, we deliberately increase the probability of token predictions $x_{t}$ that fit the token context $x_{<t}$ poorly, thereby simulating the failure case of unreliably estimated tokens reported for conventional sampling (see \cref{sec:background}), where truncation techniques are expected to provide better results.
We empirically set $r$ to the minimal value that leads to an audible degradation of the generated sequences. This resulted in $r=1.5$. 
The NLL of the test data under the temperature-degraded model is measured to be $NLL=0.31$, which is an increase of $0.01$ compared to the well-calibrated model.

Secondly, we consider an unbiased degradation where we perturb the network weights by adding a small amount of Gaussian noise. More specifically, for every weight matrix $W$ of the well-calibrated model, we obtain a degraded weight matrix $W^{\prime}$ by adding noise $z_{W}$ to $W$
\begin{equation}
    W^{\prime} = W + k z_{W},
\end{equation}
where $z_{W} \sim \mathcal{N} (0, \text{std}(W))$ and $k$ is a constant.
We sample the noise vector once and keep it fixed for all our experiments.
We empirically set $k$ to be the minimal value where sample degradations are audible, which results in $k=0.175$. The NLL of the test data under the resulting model is measured as $NLL=0.36$, which is an increase of $0.06$. 

\subsection{Generation}\label{sec:generation}
When generating sequences with the learned models, for all models, we perform conventional sampling, nucleus sampling and typical sampling as described in \cref{sec:methods}. We sample until either the end-of-sequence token is encountered or a maximum length is reached. Due to computing limitations, we fix the maximum sequence length to the 80\%-quantile of the dataset song-length distribution.
We keep both sequences which terminate with the end-of-sequence token and sequences with the maximum length reached in our sample sets.
\subsection{Objective Evaluation}
The objective evaluations are performed by calculating different statistics from the generated sequences and comparing the results between different (non-)degradations, sampling types and with the original reference data.
\subsubsection{Surprisal}
We are interested in the degree of surprisal of the samples generated with the different sampling methods. Similar to \cite{meyer,idyom,tismir}, we measure surprisal using the IC of events. As we do not have access to the data distribution, we interpret the well-calibrated model to be an oracle that approximates the data distribution. We then use the well-calibrated model to measure the mean IC of all events from a specific sampling method and model.

\subsubsection{Structural Consistency}
We measure structural consistency by investigating the self-similarities of the generated pieces. Similar to \cite{museformer}, we compute a self-similarity distribution from samples of a given sampling method and contrast it with the similarity distribution calculated from real data.
To do so, we first compute the similarity between bar pairs separated by measure lags of size $t$. This is done for each tune $x$ in sample sets $D$  according to 
\begin{equation}
    l_{i,i+t}^{x} = \frac{\left\vert N\left(i\right)\cap N\left(i+t\right) \right\vert}{\left\vert N\left(i\right)\cup N\left(i+t\right) \right\vert}, 
\end{equation}
where the set of notes in the $i$-th bar is denoted as $N(i)$, and two notes are deemed equal if their pitches, durations, and onset positions within their respective bars are identical. The similarity score $l^{x}_{i,j}$ between any two bars ranges from 0.0 to 1.0, with a score of 1.0 indicating that the two bars are identical.
After computing the similarity for all possible lags in each tune of a sample set $D$, we calculate the average similarity scores of that sample set by
\begin{equation}
    L^{D}_{t} = \frac{1}{\vert D \vert}\left(\sum_{x \in D} \sum_{j=i+t} l^{x}_{i,j}\right).
    \label{eq:sim}
\end{equation}
Note that \cref{eq:sim} does not define a probability distribution and does not, in general, sum to one.
For each dataset, we then calculate an overall self-similarity score
\begin{equation}
    \label{eq:ss}
    SS(D) = \frac{1}{T} \sum_{t=1}^{T}  L^{D}_{t},
\end{equation}
where $T$ is the maximum bar lag considered.
$SS(D)$ captures both short-term self-similarities, e.g., repetitions or variations of motives, and long-term self-similarities, e.g., repetitions or variations of musical segments.
Similar to \cite{museformer}, we also consider the deviation of a sample set's similarity distribution $L^{D}_{t}$ to the dataset's similarity distribution $L_{t}$ given by
\begin{equation}
    \label{eq:se}
    SE(D) = \frac{1}{T} \sum_{t=1}^{T} \left\vert L_{t} - L^{D}_{t} \right\vert.
\end{equation}
We interpret this deviation as a measure of how closely the self-similarities of tunes generated with the different sampling techniques follow the self-similarities of tunes found in the dataset. We set $T = 38$ in our experiments (i.e., the smallest maximum number of bars generated by any method).

\subsubsection{Tonal Consistency}
We are furthermore interested in the tonality coherence of samples generated with the sampling methods. Specifically, we investigate the scale consistency \cite{scalecons}, i.e., the maximum percentage of notes fitting a diatonic scale. The scale consistency is therefore calculated by
\begin{equation}
        \max_{scale} \frac{\#pitch\_in\_scale(x,scale)}{\#pitches(x)}.
        \label{eq:scalecons}
\end{equation}
A scale consistency value of 1.0 indicates that all pitches are within a single scale, whereas lower values indicate more complex harmonic structures.

\subsection{User Study}
In addition to the objective evaluations described above, we also perform a user study to gather subjective evaluations of the tunes' musical quality, structural properties and complexity. For that, we hosted a website consisting of two pages. The first page explains the purpose of the study, specifically that it aims to evaluate sampling techniques for neural network music generation. Furthermore, the users are instructed to rate the respective tunes using the attributes \emph{overall quality}, \emph{short-term structure}, \emph{long-term structure} and \emph{complexity} using a 5-point Likert scale. The users are also asked to use appropriate headphones or loudspeakers and to announce their level of musical expertise with choices \{\emph{Beginner}, \emph{Intermediate}, \emph{Expert}\}. On the second page, a list of $10$ audio widgets is displayed, one for each tune.
Below each widget, the Likert scales for the $4$ different attributes (as described above) are provided for voting. In addition, the users can click on a ``sheet link'' that opens a window displaying the tune in staff notation. The $10$ tunes for every user constitute the Cartesian product of all three sampling methods (i.e., \emph{conventional}, \emph{nucleus}, \emph{typical}) and all three model modes (i.e., \emph{well-calibrated}, \emph{temperature degradation}, \emph{noise-degradation}) plus a reference tune. It is ensured that every user obtains unique tunes sampled randomly from a set of 500 instances for each of the $10$ types, presented in a random order. To prevent biases, every user is allowed to perform the study only once.

\section{Results and Discussion}
In this section, we present the results of the experiments described in Section \ref{sec:exp}. For the figures and tables, we use the abbreviations WELL, NOISE and TEMP for the well-calibrated, noise-degraded and temperature-degraded models, respectively. To these abbreviations, we append CONV, NUCL and TYP for conventional sampling, nucleus sampling and typical sampling correspondingly.

\subsection{Objective Evaluation}
In the following section, we analyse and discuss the results of our objective and subjective evaluations.
\subsubsection{Surprisal}
\begin{figure}
    \centering
    \includegraphics[scale=.5]{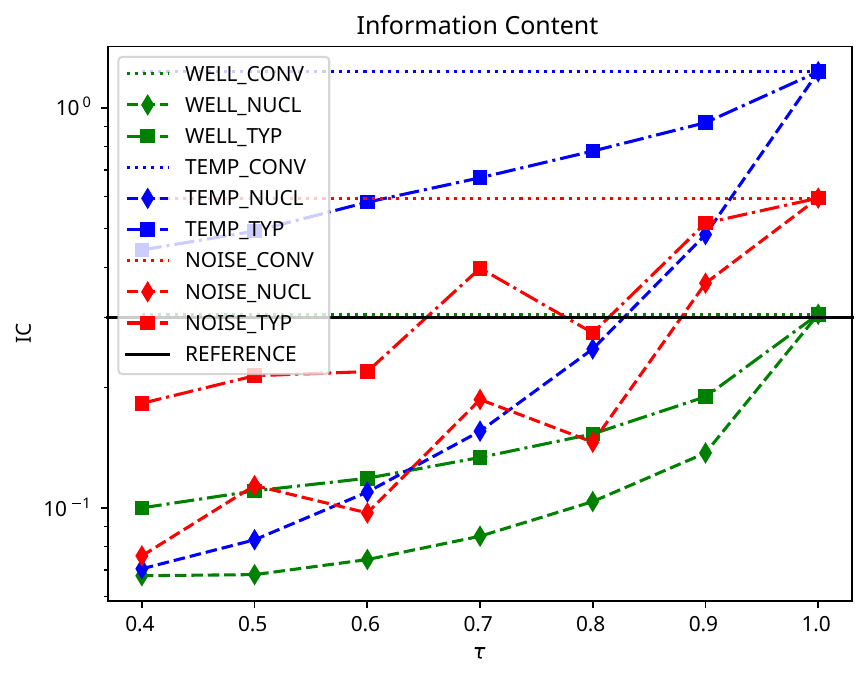}
    \caption{Information content of generated data using different sampling strategies and $\tau$ values under the well-calibrated model.}
    \label{fig:ic}
\end{figure}

We report the results of the IC estimation in \Cref{fig:ic} for the truncation degrees $\tau=0.4, ..., 1.0$. The samples from the well-calibrated model have the lowest IC and the IC of samples from the temperature-degraded model is higher than the IC of samples from the noise-degraded model. For both nucleus and typical sampling, the IC decreases with decreasing $\tau$. For typical sampling in particular, this suggests that relatively more high information than low information tokens are pruned, similar to what is found in \cite{bjarelattner}. For most degradation scenarios and sampling methods, a $\tau$ value between $0.8$ and $0.9$ is shown to recover the original data distribution best.

\subsubsection{Structural Consistency}
\begin{figure*}
    \centering
     \begin{subfigure}[t]{0.33\textwidth}
    \includegraphics[scale=.39]{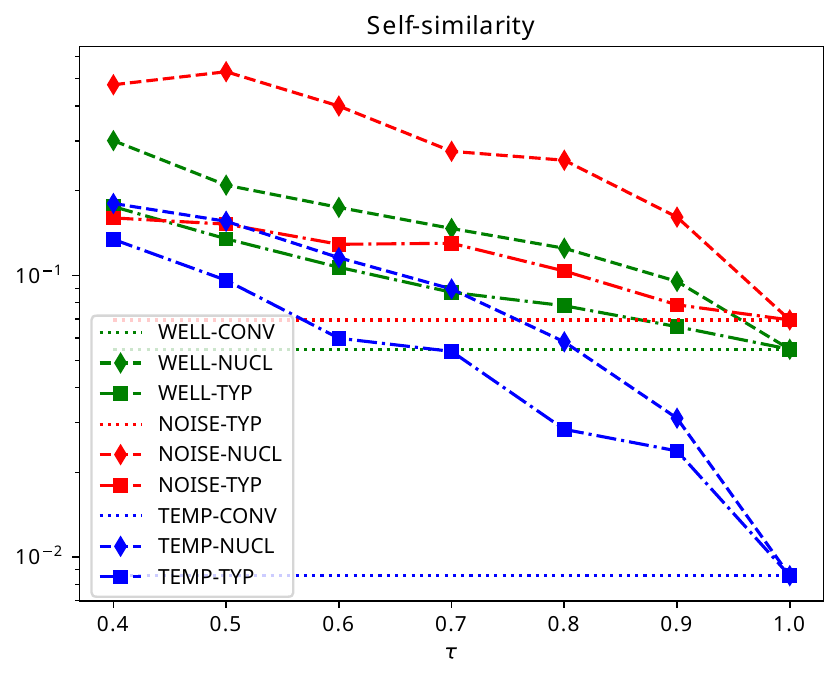}
    \caption{}
    \label{fig:ss}
    \end{subfigure}
    \begin{subfigure}[t]{0.33\textwidth}
    \centering
    \includegraphics[scale=.39]{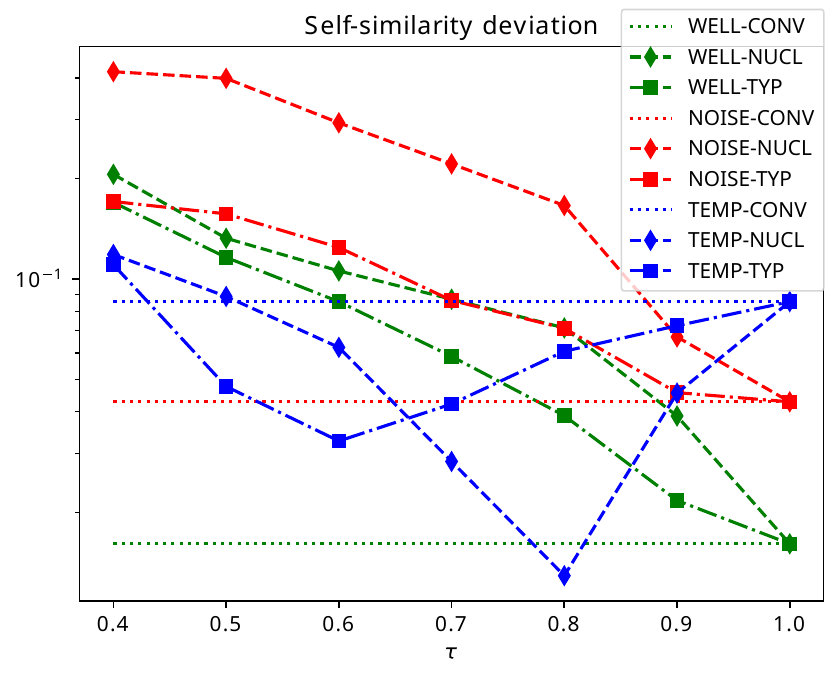}
    \caption{}
     \label{fig:se}
    \end{subfigure}
    \begin{subfigure}[t]{0.33\textwidth}
    \centering
    \includegraphics[scale=.39]{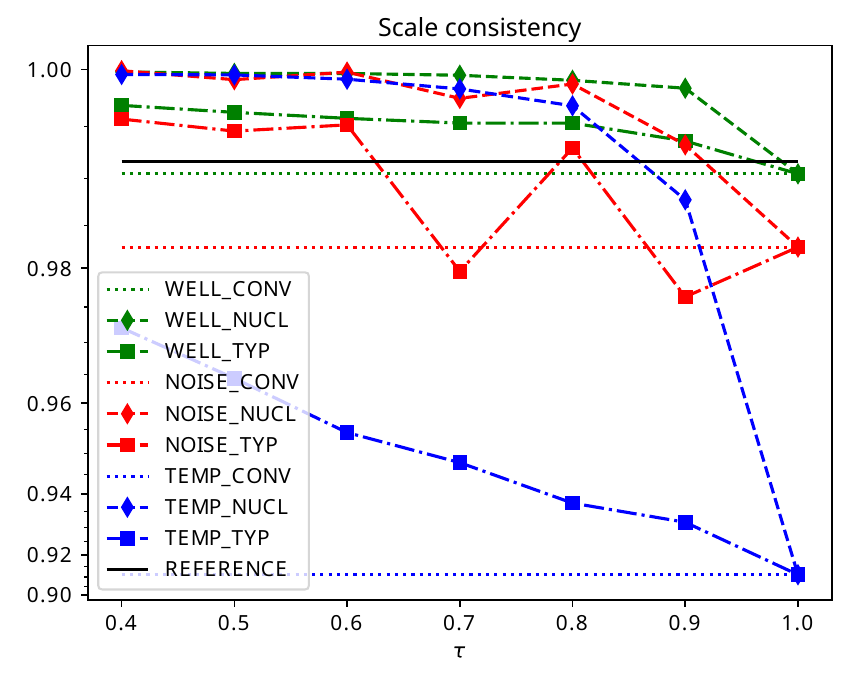}
    \caption{
        }
    \label{fig:scalecons}
    \end{subfigure}
    \caption{Structural and tonal consistency for different model degradations, sampling strategies and $\tau$ values. In (a) the self-similarity of sample sets generated with different sampling techniques is shown. Higher values indicate a higher degree of self-similarities. In (b) the deviation of the generated samples' self-similarities to the self-similarity and the data reference distribution is shown.
    A deviation of 0 indicates that the self-similarity of a sample set fits the reference distribution exactly. In (c) the scale consistency of different sample strategies and the reference dataset is shown. }
\end{figure*}

We compute the self-similarity (see \Cref{eq:ss}) for all models and sampling techniques and show the result in \Cref{fig:ss}.
Similarly, we plot the self-similarity deviation (see \Cref{eq:se}) in \Cref{fig:se}.
From \Cref{fig:ss}, we find that the overall self-similarity of samples produced with typical and nucleus sampling increases as $\tau$ decreases. This holds for both degraded models and the well-calibrated model. However, we find that the increase in self-similarity is more moderate for samples generated with typical sampling than those of nucleus sampling, indicating that the removal of highly probable tokens keeps the self-similarity at more moderate levels.  
In the temperature degradation scenario, we find that moderate levels of truncation lower the self-similarity deviation for the temperature-degraded model and thereby counteract the temperature degradation (with an optimal $\tau$ of $0.8$ and $0.6$ for nucleus sampling and typical sampling, respectively). In fact, in this scenario, the self-similarity of samples generated with nucleus and typical sampling follows the self-similarity of the reference distribution closer than samples generated with ordinary sampling for most tested truncation strengths.
This is not the case for the unbiased noise degradation, where the self-similarity increases with higher truncation strengths, increasing also the deviation from the reference statistics.

\subsubsection{Tonal Consistency}
We inspect the tonal consistency by calculating the scale consistency (see \Cref{eq:scalecons}) and report the results in \Cref{fig:scalecons}.
For both nucleus and typical sampling, we find that samples generated with low values of $\tau$ lead to a higher degree of scale consistency. Furthermore, we find for any given $\tau$ that generations from typical sampling have lower scale consistency than samples generated with nucleus sampling. Especially when considering temperature degradation, the scale consistency of nucleus sampling is almost at the level of the reference distribution at $\tau=0.9$, whereas typical sampling stays low even at high levels of $\tau$. An important observation is that (with the exception of typical sampling in the temperature degradation scenario) there is an optimal $\tau$ for both truncation techniques that leads to a recovery of the dataset's scale consistency statistic in both degradation scenarios.
\\

Similar to the findings in \cite{nucleus} for natural language, our objective evaluations in the high-temperature scenario indicate that the musical statistics of the samples generated with truncation techniques more closely match the statistics of samples from the reference distribution. This finding implies that truncation sampling techniques can be applied to music generative language models, similar to their application in natural language. This can help remove tokens with unreliable probability estimates that do not fit the musical context well. This approach may have implications for more complex datasets and limited resources, where obtaining a well-calibrated model can be challenging.

\subsection{User Study}
\begin{table}
\centering
\small
\begin{tabular}{@{}lllll@{}}
\toprule
   Method                             & QULT   & ST\_STR   & LT\_STR   & CPLX   \\
\midrule
REFERENCE                        & \textbf{3.7}±1.0           & 3.8±1.0      & \textbf{3.7}±1.1      & 3.6±0.8      \\
WELL\_CONV            & 3.2±1.1           & 3.7±0.9      & 3.5±1.2      & 3.3±1.0      \\
WELL\_NUCL       & 3.6±1.1           & \textbf{3.9}±1.1      & \textbf{3.7}±1.1      & 2.8±1.0      \\
WELL\_TYP       & 3.4±1.2           & 3.6±0.9      & \textbf{3.7}±1.0      & 3.3±1.0      \\
NOISE\_CONV      & 2.7±1.0           & 3.2±0.9      & 3.0±1.0      & 2.8±0.9      \\
NOISE\_NUCL & 2.6±1.3           & 3.2±1.4      & 2.8±1.5      & 2.5±1.2      \\
NOISE\_TYP & 2.7±1.1           & 3.2±1.1      & 3.1±1.2      & 2.4±1.0      \\
TEMP\_CONV            & 2.1±1.3           & 2.7±1.1      & 2.1±1.1      & \textbf{3.7}±1.0      \\
TEMP\_NUCL       & 3.4±1.2           & 3.6±0.9      & 3.4±1.3      & 3.4±1.1      \\
TEMP\_TYP       & 2.2±1.1           & 2.7±0.9      & 2.4±1.0      & 3.3±0.8      \\
\bottomrule
\end{tabular}
\caption{Results showing the mean-opinion scores of the user study $\pm$ the standard deviation. QULT denotes the overall quality estimation, ST\_STR the perceived short-term structure, LT\_STR the perceived long-term structure and CPLX the perceived complexity of the rated samples.}
\label{tab:subjective}

\end{table}

The user study was performed by $38$ participants who,  according to their self-assessment can be divided into $8$ beginners, $18$ intermediate and $12$ musical experts. The presented melodies (except the \emph{reference}) are generated as described in Section \ref{sec:generation}, with $\tau = 0.8$ for both, nucleus and typical sampling.
Table \ref{tab:subjective} shows the user study results. As there is a high variance for all ratings, we performed for all attributes a Welch's t-test between all $m=10$ tune types. Using a desired significance level of $\alpha = 0.05$, the corresponding Bonferroni correction to the multiple comparisons problem gives a significance level of $\frac{\alpha}{\frac{1}{2}m(m-1)}=\frac{0.05}{45}=0.001$.
We can see in the first column that the human-composed reference tracks have the highest quality scores on average and that the perceived quality of the tunes tends to degrade for the noise- and temperature degradation cases as expected. The t-test shows that the users' preference for REFERENCE is significant compared to all samples of the under-calibrated models (with $p < 1 \times 10^{-4}$), except for TEMP\_NUCL with $p=0.37$. This shows that nucleus sampling can potentially improve the sample quality of low-confidence models, while typical sampling is not able to recover any degradations. 
Furthermore, we find that WELL\_CONV, WELL\_NUCL and WELL\_TYP differ in QULT with $p=0.07,0.67$ and $0.37$ respectively compared to REFERENCE. This provides some evidence that nucleus and typical sampling improves the sampling quality of well-calibrated models, but this effect is not significant.
While nucleus sampling performs well in the temperature-degraded model, we observe some (non-significant) evidence of a lower complexity than conventional and typical sampling in the well-calibrated model (with $p=0.023$ and $p=0.044$, respectively).
Typical sampling (with $\tau=0.8$) does not cause significant differences from conventional sampling. 
As the $p$-value between NOISE\_TYP and NOISE\_CONV is also low (but not significant, with $p=0.06$), there is some evidence that typical sampling slightly reduces the complexity of outputs from under-calibrated models. This could be explained by typical sampling pruning the higher and lower probability events, overall reducing the possible number of events to be sampled.
The well-calibrated model performs well with all sampling techniques (no significant differences to REFERENCE), with only some non-significant evidence for lower complexity with nucleus sampling.

\section{Conclusion}
We investigated the effect of distribution truncation sampling techniques on the musical qualities of information content, self-similarity, scale consistency and complexity of samples generated under different degradation scenarios. Our objective evaluations show that a higher truncation strength leads to increased self-similarity and tonal consistency. This trend is more pronounced for samples generated with nucleus sampling compared to samples generated with typical sampling. For a well-calibrated model, we show that the increase in self-similarity and scale consistency leads to an increase in deviations of these metrics from the reference distribution. However, for under-calibrated models, we showed that the deviations from the original data statistics could often be reduced with the correct truncation strategy and carefully selected truncation levels (where a $\tau$ between $0.8$ and $0.9$ seems to be good trade-off value over all experiments). While nucleus sampling carries the risk to reduce complexity of the outputs, this trend could not be observed with typical sampling.
\newpage
\section{Acknowledgments}
The work leading to these results was conducted in a collaboration between JKU and Sony Computer Science Laboratories Paris under a research agreement. GW's work is supported by the European Research Council (ERC) under the European Union’s Horizon 2020 research and innovation programme, grant agreement 101019375 (\textit{“Whither Music?”}).
\bibliography{ISMIRtemplate}
\end{document}